\documentclass[conference]{IEEEtran}

\usepackage{amsmath}
\usepackage{amssymb}
\usepackage{listings} 
\usepackage{url}
\usepackage{graphicx}
\usepackage{nameref}
\usepackage{multirow}

\usepackage[inline]{enumitem}
\usepackage[flushleft]{threeparttable}
\usepackage{array}
\usepackage{multicol}
\usepackage{graphicx}
\usepackage[nospace,noadjust]{cite}

\begin{document}

\title{Using Temporal and Semantic Developer-Level Information to Predict Maintenance Activity Profiles}

\author{\IEEEauthorblockN{Stanislav Levin}
\IEEEauthorblockA{The Blavatnik School of Computer Science\\
 Tel Aviv University\\
Tel-Aviv, Israel\\
stanisl@post.tau.ac.il}
\and
\IEEEauthorblockN{Amiram Yehudai}
\IEEEauthorblockA{The Blavatnik School of Computer Science\\
 Tel Aviv University\\
Tel-Aviv, Israel\\
amiramy@tau.ac.il}
}

\maketitle

\begin{abstract}
Predictive models for software projects' characteristics have been traditionally based on project-level metrics, employing only little developer-level information, or none at all.
In this work we suggest novel metrics that capture temporal and semantic developer-level information collected on a per developer basis. To address the scalability challenges involved in computing these metrics for each and every developer for a large number of source code repositories, we have built a designated repository mining platform. This platform was used to create a metrics dataset based on processing nearly 1000 highly popular open source GitHub repositories, consisting of 147 million LOC, and maintained by 30,000 developers. The computed metrics were then employed to predict the corrective, perfective, and adaptive maintenance activity profiles identified in previous works. Our results show both strong correlation and promising predictive power with $R^2$ values of $0.83$, $0.64$, and $0.75$. We also show how these results may help project managers to detect anomalies in the development process and to build better development teams. In addition, the platform we built has the potential to yield further predictive models leveraging developer-level metrics at scale. 

\end{abstract}


\section{Introduction}

Forecasting maintenance activities performed in a source code repository could help practitioners reduce uncertainty and improve cost-effectiveness \cite{swanson1976dimensions} by planning ahead and pre-allocating resources towards source code maintenance. 
Maintenance activity profiles of software projects have been therefore a subject of research in numerous works \cite{swanson1976dimensions,mockus2000identifying, meyers1988, lientz1978characteristics}. 
In this work we adopt the maintenance activity categories \textbf{corrective}, \textbf{perfective}, and \textbf{adaptive} as defined by Mockus et al. \cite{mockus2000identifying}:
\begin{itemize}\label{sec:changeClassification}
    \item Corrective: fault fixing. 
    \item Perfective: code structure / system design improvements.
    \item Adaptive: new feature introduction.
\end{itemize}
and put forth previously unexplored \textbf{temporal and semantic developer-level metrics}, which we then utilize to study the corrective, perfective and adaptive maintenance activity profiles on a developer-level granularity. 

We seek to gain better understating of how personal characteristics such as commit patterns, commit frequencies, project join and departures dates, etc., impact the maintenance activities of individual developers, as well as the projects they work on. 
Moreover, given a software project, we consider its maintenance activity profiles as an aggregation of the maintenance activity profiles of all developers' that have taken part in its development. Therefore, predictions made on a developer-level (i.e., for all individual developers in the project), can be used to reason about, and derive project-level predictions. 

Studying developer-level impact requires sufficient and sufficiently fine-grained data, as well as the computational power to process it. This work is therefore driven by two main factors that have been trending up for the past decade: 
\begin{itemize}
  \item Big Code \cite{bigCode}: the availability of large source code corpora via open source.
  \item Big Data \cite{diebold2012origin} ecosystem: the availability of tools capable of processing extremely large data volumes.
\end{itemize}
The combination of the two has created unprecedented opportunities to collect and process an enormous volume of source code, and provide insights that were previously exponentially harder, or even impossible to obtain \cite{raychev2015predicting}.

\section{Related Work}

Maintenance activity profiles and the application thereof have been the subject of numerous works dealing with fault prediction models, commit classification, software change recommendations, and more \cite{schach2003determining, lientz1978characteristics, johnson1988designing, mockus2000identifying, hassan2009predicting, zimmermann2005mining}. While the precise distribution of maintenance activities is inconclusive \cite{schach2003determining}, their classification into the corrective, perfective and adaptive categories has been a common practice. 
Our work therefore seeks to explore how these maintenance activity categories relate to the semantic and temporal developer-level metrics we define, and whether such metrics can be used to build effective predictive models for the corrective, perfective and adaptive maintenance activity profiles.

\section{Research Question}

\section*{\textbf{\textit{RQ: How do temporal and semantic developer-level metrics relate to developers maintenance activity profiles?}}\label{rq2}}

To the best of our knowledge, this work is the first to explore temporal and semantic \textit{developer-level} metrics (see Table \ref{tab:developerMetrics}), such as the number of distinct semantic changes performed by each developer, the mean number of distinct semantic changes in a given developer's commits, mean time between commits and others. Moreover, we explore these metrics at large scale, and analyze nearly 1000 different repositories that consist of dozens of millions of commits in total. Our large scale study was conducted using a VCS mining platform we have built to enable large scale analysis of version control systems (VCS). Our platform leverages Spark \cite{zaharia2010spark}, an industrial cutting edge data processing framework.

We used the GQM approach \cite{basili1992software} to derive the questions, and then the metrics that are needed in order to answer the
research question in a measurable way. The developer-level metrics we measured are listed in Table \ref{tab:developerMetrics}.

\begin{table}[!htbp] 
\caption{Developer-level\tnote{1} metrics}
\label{tab:developerMetrics} 
    \begin{threeparttable}
        \begin{tabular}{m{3.7cm}|m{4.3cm}}
         \hline
            $\operatorname{Commits}_{repo}$ & \multirow{2}{*}{\parbox{4.3cm}{\center See Table \ref{lst:versatilityFormula}}} \\\cline{1-1}
            $\operatorname{Muse}_{repo}$ &  \\\cline{1-1}
            $\operatorname{DeveloperVersatility}_{repo}$ & \\\cline{1-1}
            $\operatorname{MeanCommitV}_{repo}$ & \\\cline{1-1}
            $\operatorname{VersatilityLevel}_{repo}$ &  \\\cline{1-1}
        \hline
           \parbox{2.8cm}{$\operatorname{ContribStartRel}_{repo}$} & Developer's join date (first recorded commit) expressed as the number of days since project's first observed commit  \\
        \hline
             \parbox{2.8cm}{$\operatorname{ContribDuration}_{repo}$} & The duration in days, of the period between the developer's first and last commit dates. \\
        \hline
            $\operatorname{MTBC}_{repo}$ &  \underline{\textbf{M}}ean \underline{\textbf{t}}ime \underline{\textbf{b}}etween developer's \underline{\textbf{c}}ommits, in days. \\
        \hline
            
        \hline
        \end{tabular}
        \begin{tablenotes}
          \footnotesize
          \item[*] Developers who have committed changes (contributed) to multiple source code repositories are considered as if they were different individuals.
        \end{tablenotes}
    \end{threeparttable}
\end{table}

\section{Background}
\label{sec:dataCollection}

Fluri et al. \cite{fluri2006classifying} put forth  a taxonomy of semantic source code changes for object-oriented programming languages (OOPLs), and Java in particular. This taxonomy consists of 47 different change types, such as \textit{statement\_delete}, \textit{statement\_insert},
\textit{statement\_update}, \textit{removed\_class}, \textit{additional\_class}, \textit{return\_type\_change} and so on. The computation of these change types from source code files was later implemented by a tool named "ChangeDistiller" \cite{gall2009change}.

We embrace this taxonomy and define the notion of versatility based measures for developers and commits as defined in Table \ref{lst:versatilityFormula}.

\begin{table*}
  \caption{Versatility based measures for developers and commits}
  \label{lst:versatilityFormula}
  \centering
  \begin{tabular}{|l|}
  \hline
Given a source code repository $repo$ and a developer $dev$ that has committed code to $repo$: \\
  
$\boldsymbol{\operatorname{Commits}_{repo}(dev)} = \{\text{all commits by developer $dev$ to repository $repo$}\}$ \\ $\boldsymbol{\operatorname{CommitVSet}_{repo}(commit)}:= \{ \text{all semantic source code changes in } commit \}$ \\

$\boldsymbol{\operatorname{DeveloperVSet}_{repo}(dev)} := \bigcup\limits_{commit \in \operatorname{Commits}_{repo}(dev) } \operatorname{CommitVSet}_{repo}(commit)$ \\
$\boldsymbol{\operatorname{CommitVersatility}_{repo}(commit)} :=  \mid \operatorname{CommitVSet}_{repo}(commit) \mid $ \\
$\boldsymbol{\operatorname{DeveloperVersatility}_{repo}(dev)} := \mid \operatorname{DeveloperVSet}_{repo}(dev) \mid$ \\
$\boldsymbol{\operatorname{Muse}_{repo}(dev)} := \max\limits_{commit \in \operatorname{Commits}_{repo}(dev) } \operatorname{CommitVersatility}_{repo}(commit) $ \\
$\boldsymbol{\overline{\operatorname{CommitVersatility}}_{repo}(dev)} := 
(\sum\limits_{commit \in \operatorname{Commits}_{repo}(dev) } \operatorname{CommitVersatilisy}_{repo}(commit) ) * \frac{1}{\mid \operatorname{Commits}_{repo}(dev)\mid } $ \\
$\boldsymbol{\operatorname{VersatilityLevel}_{repo}(dev)} := \mid \bigcup\limits_{commit \in \operatorname{Commits}_{repo}(dev) } \{S | S = \operatorname{CommitVSet}_{repo}(commit) \} \mid$ \\
Note that: $ \operatorname{DeveloperVersatility}_{repo}(dev) \geq
\operatorname{Muse}_{repo}(dev)$ \\
\hline
  \end{tabular}
\end{table*}

\noindent For example, if developer $Alice$ performed 2 commits:
\begin{center}
 \begin{tabular}{l|l}
 $commit_1$ & 2 x \textit{stmt\_insert}, 1 x \textit{stmt\_update}  \\
 \hline
 $commit_2$ & 3 x \textit{stmt\_delete}  \\
  \end{tabular}
\end{center}

\noindent
\resizebox{\linewidth}{!}{%
\renewcommand{\arraystretch}{1.5}
\noindent
\begin{tabular}{@{}l}
$\operatorname{CommitVSet}_{repo}(commit_1) = \{ \text{stmt\_insert}, \text{stmt\_update}\}$ \\
$\operatorname{CommitVSet}_{repo}(commit_2) = \{ \text{stmt\_delete}\}$ \\
$\operatorname{DeveloperVSet}_{repo}(Alice) = \{ \text{stmt\_insert}, \text{stmt\_update},\text{stmt\_delete}\}$ \\
$\operatorname{CommitVersatility}_{repo}(commit_1)  = 2$ \\
$\operatorname{CommitVersatility}_{repo}(commit_2)  = 1$ \\
$\overline{\operatorname{CommitVersatility}}_{repo}(Alice)  = \frac{2+1}{2} = 1.5$ \\
$\operatorname{DeveloperVersatility}_{repo}(Alice)  = \mid \operatorname{DeveloperVSet}_{repo}(Alice) \mid = 3$ \\
$\operatorname{Muse}_{repo}(Alice)  = \max  \{2,1\} = 2$ \\
$\operatorname{VersatilityLevel}_{repo}(dev) = \mid \{ \{ \text{stmt\_insert}, \text{stmt\_update}\},\{ \text{stmt\_delete}\} \} \mid  = 2 $ \\
\end{tabular}%
}
\\

The intuition behind these versatility based measures is capturing the number of different fine-grained semantic change types present in commits made by individual developers. 

\section{Metrics Computation}\label{eval}

To provide reproducible results, we sought to base our empirical study on publicly (and freely) available data. We chose GitHub as the data source for this work due to its prevalence among the repository hosting services. Candidate repositories were selected according to the following criteria:
\begin{enumerate}
    \item used the Java programming language
    \item had more than 100 stars (i.e. more than 100 users had "liked" these repositories)
    \item had more than 60 forks (i.e., more than 60 users had "copied" these repositories for their own use)
    \item had their code updated since 2016-01-01 (i.e., these repositories were active)
    \item were created before 2015-01-01  (i.e., these repositories had been around for at least $\sim$1.5 years)
    \item had size over 2MB (i.e. these repositories were beefy)
\end{enumerate}
The result set consists of 1000 repositories, the maximum limit as stated by GitHub's documentation \cite{gitHubSerchLimit}.
Due to various technical reasons our final dataset consisted of 979 unique Git repositories, where the average number of developers per project was 32, and the average project age was 4.2 years. These repositories were then cloned and processed by our VCS mining platform

First, to distill semantic source code changes as per the taxonomy defined by Fluri et al. (see also section \ref{sec:dataCollection}), our VCS mining platform repeatedly applied the ChangeDistiller (\cite{gall2009change, fluri2008discovering, martinez2013automatically}) on every two consecutive revisions of every Java file in every repository in our result set. This stage yielded 30 million semantic source code change instances.
Then, all the semantic source code changes were aggregated using the key (developer-id, repository-id), forming data bins from which the temporal and semantic developer-level metrics (see Table \ref{tab:developerMetrics}) were computed. To classify developer's commits into the corrective, perfective, and adaptive categories, our VCS mining platform used methods similar to \cite{mockus2000identifying, fischer2003populating, sliwerski2005changes}, and searched for indicative keywords in the commit's comment field. Keyword matching was boosted by using common techniques such as stemming and case-folding, the keywords are listed in table \ref{tab:classificationWords}.

\begin{table}[!ht]
\caption{Keywords for classifying maintenance activities}
\label{tab:classificationWords} 
    \begin{tabular}{|p{1.2cm}|p{6.8cm}|}
    \hline
        Corrective & \textit{fix,
          esolv,
          clos,
          handl,
          issue,
          defect,
          bug,
          problem,
          ticket} \\
    \hline
        Perfective & \textit{refactor,
          re-factor,
          reimplement,
          re-implement,
          design,
          replac,
          modify,
          updat,
          upgrad,
          cleanup,
          clean-up} \\
    \hline
        Adaptive &  \textit{add,
          new,
          introduc,
          implement,
          implemented,
          extend,
          feature,
          support} \\
    \hline
    \end{tabular}
    \vspace{-1em}
\end{table}

\section{Results}\label{sec:results}

We use generalized regression modeling (GLM) \cite{mccullagh1989generalized} in the R statistical environment \cite{R} to explore our dataset and build predictive models. 
The predictive models were trained on randomly chosen $90\%$ of the repositories in the dataset, while the remaining $10\%$ were used for validation and measuring goodness of fit.

In the rest of this section we present the predictive models (see Table \ref{tab:regressionModels}) for the developer-level corrective, perfective and adaptive maintenance activity profiles. 
\noindent To predict the profile of a maintenance activity category $\operatorname{MA}_c \in \{\operatorname{Corrective}, \operatorname{Perfective}, \operatorname{Adaptive} \}$ for a developer $dev$ the following formula was used:
\noindent\resizebox{\linewidth}{!}{
$\operatorname{Profile}_{\operatorname{MA}_c}(dev) = 
\operatorname{Const_{\operatorname{MA}_c}} + \sum\limits_{P_i \in model_{\operatorname{MA}_C}}\operatorname{coeff}_{\operatorname{MA}_c}^{P_i} * P_i(dev)$}
where $model_{\operatorname{MA}_c}$ is the regression model for $\operatorname{MA}_c$,
$\operatorname{Const_{MA_c}}$ is the constant in $model_{\operatorname{MA}_c}$, and $\operatorname{coeff}_{\operatorname{MA}_c}^{P_i}$ is the coefficient of predictor $P_i$ in $model_{\operatorname{MA}_c}$ as specified in table \ref{tab:regressionModels}. $P_i(dev)$ is the value of predictor $P_i$ for a given developer $dev$.

\begin{figure}
    \caption{Prediction graphs for corrective, perfective and adaptive (left to right) maintenance activity profiles for 150 developers randomly selected form the test dataset. Actual values in blue, predicted ones are in red.}
    \label{fig:predictionPlots}
    \includegraphics[width=0.5\textwidth,height=3cm]{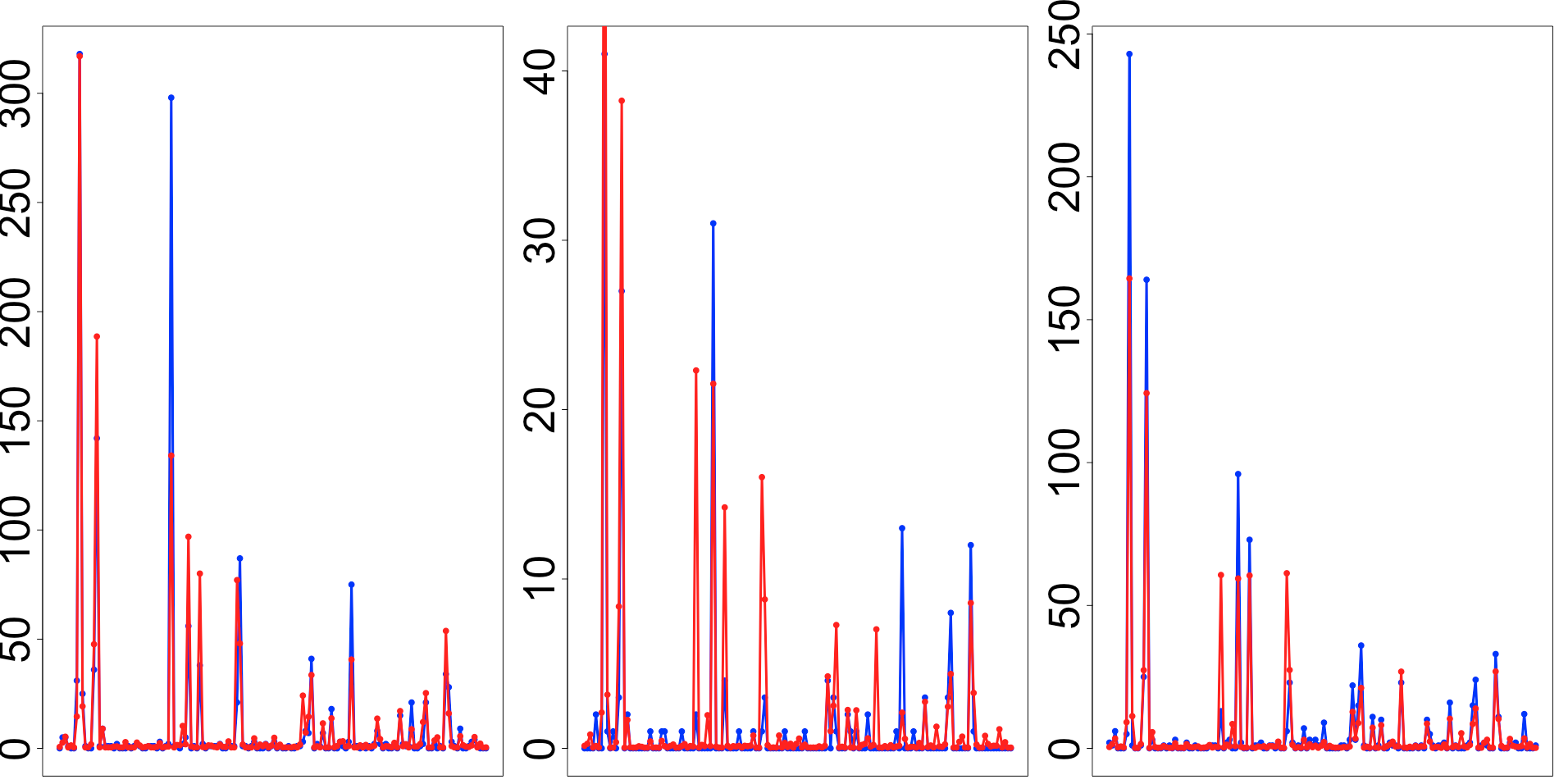}
\end{figure}

All predictors were log transformed to alleviate skewed data, a common practise when dealing with software metrics \cite{shihab2012exploration, camargo2009towards}. The standard error is specified in parenthesis underneath the estimate for each predictor.
Figure \ref{fig:predictionPlots} presents predictions results for 150 developers randomly selected from the test dataset. 

For each maintenance activity category $\operatorname{MA}_c$, the X axis is a running developer-id and the Y axis is the number of the developer's commits of category $\operatorname{MA}_c$. For each developer, we plot in blue the number of her commits classified as $\operatorname{MA}_c$ by the commit message classification algorithm (see section \ref{eval}), and overlay it in red with the number of commits predicted to be of category $\operatorname{MA}_c$ by the GLM (see Table \ref{tab:regressionModels}).

For all three maintenance activity profiles, the vast majority of both temporal and versatility based developer metrics were statistically significant with \textit{p-value} $<0.01$. Metrics that were not statistically significant were excluded from the predictive models (represented by the empty cells in table \ref{tab:regressionModels}) to improve prediction quality.

\begin{table}
\newcolumntype{P}[1]{>{\centering\arraybackslash}p{#1}}
\centering 
  \caption{GLM for developer-level maintenance activity} 
  \label{tab:regressionModels} 
\begin{tabular}{m{0.3cm}m{3.5cm}P{1cm}P{1cm}P{1cm}} 
 & & \multicolumn{3}{c}{\textit{Predicted profile:}} \\ 
\cline{3-5} 
\\[-1.8ex] \multicolumn{2}{c}{\textbf{Predictor}} &  \textbf{\footnotesize Corrective} \tiny(1) & \textbf{\footnotesize Perfective} \tiny(2) & \textbf{\footnotesize Adaptive} \tiny(3) \\ 
\cline{2-5}  \\[-1.8ex] 
 \tiny$(P_1)$ & $log(\operatorname{Commits}_{repo})$ & 0.797 & 0.572 & 0.503 \\ 
 & & (0.010) & (0.020) & (0.015) \\ 
  \cline{2-5} \\[-2ex] 
 \tiny$(P_2)$ &$log(\operatorname{Muse}_{repo})$ & 0.171 & $-$0.288 & $-$0.135 \\ 
  & & (0.010) & (0.020) & (0.013) \\ 
  \cline{2-5}\\[-2ex] 
 \tiny$(P_3)$ & $log(\operatorname{MTBC}_{repo})$ & 0.012 & $-$0.018 &  \\ 
  & & (0.002) & (0.004) &  \\ 
  \cline{2-5}\\[-2ex] 
 \tiny$(P_4)$ & 
 \resizebox{3.8cm}{!}{%
 $log(\operatorname{ContribStartRel}_{repo}+ 0.1)$} & 0.014 &  & $-$0.021 \\ 
  & & (0.001) &  & (0.001) \\ 
  \cline{2-5}\\[-2ex] 
 \tiny$(P_5)$ & 
 \resizebox{3.8cm}{!}{%
 $log(\overline{\operatorname{CommitVersatility}}_{repo})$} & 0.028 &  & 0.033 \\ 
  & & (0.009) &  & (0.013) \\ 
  \cline{2-5}\\[-2ex] 
 \tiny$(P_6)$ & 
 \resizebox{3.8cm}{!}{%
 $log(\operatorname{ContribDuration}_{repo} + 0.1)$} & 0.030 & $-$0.050 & $-$0.018 \\ 
  & & (0.002) & (0.005) & (0.002) \\ 
  \cline{2-5}\\[-2ex] 
 \tiny$(P_7)$ & $log(\operatorname{DeveloperVersatility}_{repo})$ & $-$0.205 & 0.394 & 0.243 \\ 
  & & (0.010) & (0.025) & (0.013) \\ 
  \cline{2-5}\\[-2ex] 
 \tiny$(P_8)$ & 
 $log(\operatorname{VersatilityLevel}_{repo})$ & 0.181 & 0.483 & 0.437 \\
  & & (0.012) & (0.025) & (0.017) \\ 
  \cline{2-5}\\[-2ex] 
 & Constant & $-$0.986 & $-$3.092 & $-$1.462 \\ 
  & & (0.019) & (0.048) & (0.020) \\ 
\hline 
\hline 
& \textbf{$R^2$} & \textbf{0.832} & \textbf{0.640} & \textbf{0.759} \\ 
& Observations & 27,850 & 27,850 & 27,850 \\ 
\hline 
\end{tabular} 
\vspace{-1em}
\end{table} 

In all three models, $\operatorname{Commits}_{repo}$ (the total number of commits made by a developer to the given repository) was the most powerful predictor, and accounted for a great of deal the high $R^2$ values.

\textbf{Corrective profile} (table \ref{tab:regressionModels}, column 1):
In contrast to other predictors, \textit{Versatility${_d}$} has a negative coefficient indicating that developers with higher \textit{Versatility${_d}$} values are likely to perform less corrective commits given that other predictors remain fixed. The fact that \textit{Muse} and \textit{Versatility${_d}$} are both versatility based metrics, yet have opposite signs in this model, supports our assumption that \textit{Muse} and \textit{Versatility$_d$} capture different kinds of information. In addition, it is also evident that developers who commit less frequently (higher MTBC), join the project later, remain active for longer, and have more commits with distinct change type patterns, are likely to have a higher corrective profile (i.e., perform more corrective commits).

\textbf{Perfective profile} (see Table \ref{tab:regressionModels}, column 2):
Similarly to the corrective model, the signs of \textit{Versatility${_d}$} and \textit{Muse} are opposite, but in contrast to the former, it is now \textit{Muse} that has a negative sign. This indicates that developers with higher \textit{Muse} values are likely to perform less perfective commits given that other predictors remain fixed. Also, developers who commit more frequently (lower MTBC), and remain active for shorter time (lower ContribDuration), but have more commits with distinct change type patterns (i.e., higher values for DistinctChangeTypeSets) are likely to have a higher perfective profile (i.e., perform more perfective commits).

\textbf{Adaptive profile} (see Table \ref{tab:regressionModels}, column 3):
The adaptive model is more similar to the perfective one than to the corrective one, with \textit{Versatility} having a positive sign and \textit{Muse} a negative sign. Adaptive commits are likely to favour developers with lower Muse, who join the project earlier (lower ContribStartRel), remain active for shorter time (lower ContribDuration), have more commits with distinct change type patterns, and have a greater versatility (as defined in Table \ref{lst:versatilityFormula}).

\section{Discussion \& Applications}\label{sec:discussion}

\noindent\textbf{Identifying anomalies in development process.} 
    The manager of a large software project should aim to control and manage its maintenance activity profiles. Monitoring for unexpected spikes in maintenance activity profiles and investigating the reasons (root cause) behind them would assist managers and other stakeholders to plan ahead and identify areas that require additional resource allocation. 
    For example, lower corrective profiles could imply that developers are neglecting bug fixing. Higher corrective profiles could imply an excessive bug count. Finding the root cause in cases of significant deviations from predicted values may reveal essential issues whose removal can improve projects' health. Similarly, exceptionally well performing projects can also be a good subject for investigation in order to identify positive patterns.
    
\noindent\textbf{Improving development team's composition.}
    Building a successful software team is hardly a trivial task as it involves a delicate balance between technological and human aspects \cite{gorla2004should, guinan1998enabling}. We believe that developers' maintenance activity profiles could assist in composing a more balanced team. We conjecture that composing a team that heavily favors a particular maintenance activity over the others could lead to an unbalanced development process and adversely affect the team's ability to meet typical requirements such as developing a sustainable number of product features, adhering to quality standards, and minimizing technical debt so as to facilitate future changes.

\section{Threats to validity}
\label{sec:threatsToValidity}

\noindent\textbf{\textit{Threats to Statistical Conclusion Validity}} is the degree to which conclusions about the relationship among variables based on the data are reasonable. Our results are based on nearly 30,000 observations, and the predictors are statistically significant with \textit{p-value} $<0.01$.

\noindent\textbf{\textit{Threats to Construct Validity}} consider the relationship between theory and observation, in case the measured variables do not measure the actual factors.

\begin{itemize}
    \item \underline{Volatile Classification}. While the method we used is commonly practiced (\cite{mockus2000identifying, fischer2003populating, sliwerski2005changes}), our experiments show it may be sensitive to the choice of keywords used as indicative for the various maintenance activity types.
    \item \underline{Semantic Change Extraction}. ChangeDistiller (\cite{gall2009change,fluri2008discovering}) was used to extract semantic changes from the VCS. Unfortunately, like any other software, it may not be immune to bugs and malfunctions.
    \item \underline{Developer-level Metrics Computation}. To compute developer-level metrics we used a novel VCS mining platform we had built to support this study. While we invested great effort into testing it to ensure its proper functionality, it may not be immune to bugs and malfunctions.
\end{itemize}

\noindent\textbf{\textit{Threats to External Validity}} consider the generalization of our findings.
\begin{itemize}
    \item \underline{Programming Language Bias}. All analyzed commits were in the Java programming language. It is possible that developers who use other programming languages, have different maintenance activity patterns which have not been explored in the scope of this work.
    \item \underline{Open Source Bias}. The repositories studied in this paper were all popular open source projects from GitHub (\cite{gitHub}). It may be the case that developers' maintenance activity profiles are different in an open source environment when compared to other environments.
    \item \underline{Popularity Bias}. We intentionally selected the most popular, data rich repositories. This could limit our results to developers and repositories of high popularity, and potentially skew the perspective on characteristics found only in less popular repositories and their developers. 
\end{itemize}

\section{Conclusions and Future Work}

We have demonstrated that the developer-level metrics we have defined are statistically significant, and can be successfully used to model and predict the corrective, perfective and adaptive maintenance activity profiles with promising $R^2$ values of $0.83$, $0.64$ and $0.75$ respectively. Our work is based on studying Big Code (\cite{bigCode}), and involved processing nearly 1000 highly popular open source GitHub repositories, comprising a corpus of 147 million LOC, maintained by 30,000 developers, spread over 2.5 million revisions and 30 million individual code change instances.

We believe that considering a project's characteristics as an aggregation of the characteristics of all the developers' that have taken part in its development, and modeling the former using the latter, is a useful technique. It might, for example, assist in predicting further project characteristics such as fault potential, which has traditionally relied on project-level metrics \cite{mccabe1976complexity,menzies2007data,bell2011does,nagappan2005use,hassan2009predicting,ostrand2011predicting} or only limited developer-level metrics \cite{bell2013limited, matsumoto2010analysis, shihab2012industrial, jiang2013personalized} (e.g., numeric representations of experience, number of developers that changed a particular file or method).

In light of the promising results, we are in the process of conducting further studies that involve developer-level metrics. For example, we are working on comparing project-level and developer-level maintenance activity profiles. Preliminary results indicate that the \textit{variance} in developer-level maintenance activity profiles is much greater than it is in the project-level ones.

We believe that the combination of Big Code and fined grained developer-level metrics, can open the door to a new range of applications and empirical studies.
A particularly compelling direction is fusing more data sources such as bug tracking systems, social Q\&A sites (e.g., StackOverflow \cite{stackOverflow}) and others, to study how these types of data relate to developers' characteristics such as commit patterns, temporal activity, fault potential, etc. 
Augmented with developer surveys to validate the relations, we hope these studies could shed new light on our understanding of software development and evolution.

\section*{Acknowledgements}
We thank Ms Ilana Gelernter and the Tel Aviv University's statistical counseling service for the help with the statistical analysis, and Dr. Boris Levin for comments that greatly improved the manuscript. This research was supported by THE ISRAEL SCIENCE FOUNDATION, grant No. 476/11.

\bibliographystyle{IEEEtran}
\bibliography{bibliography}

\end{document}